\documentstyle[12pt,amstex,amssymb, amsfonts]{article}
\def\e{\mbox{e}}

\def\ib{\,\mbox{i}\,}
\def\te{\vartheta_1}
\def\tv{\vartheta_4}

\begin{document}

\title{The inversion relation and the dilute A$_{3,4,6}$
eigenspectrum.\thanks{In honour of R. J. Baxter's sixtieth birthday, this paper
was presented at the conference \emph{The Baxter Revolution in Mathematical
Physics}.}}

\author{Katherine A. Seaton\\School of Mathematical and Statistical
Sciences\\ La Trobe University\\Bundoora VIC 3083\\ Australia
\and
Murray T. Batchelor\\ Department of Mathematics\\Australian National
University\\Canberra ACT 0200\\Australia}

\date{}
\maketitle
\begin{abstract}On the basis of the result obtained by applying Baxter's exact
perturbative approach to the dilute A$_3$ model to give the E$_8$ mass
spectrum, the dilute A$_L$ inversion relation was used to predict the
eigenspectra in the $L=4$ and $L=6$ cases (corresponding to E$_7$ and E$_6$
respectively). In
calculating the next-to-leading term in the correlation lengths, or
equivalently masses, the inversion relation condition gives a surprisingly
simple result in all three cases, and for all masses.
\end{abstract}


\section{Introduction}

One model in statistical mechanics which attracts perennial interest is
the two-dimensional Ising model in a magnetic field. 
The integrable quantum field theory which
describes the (massive) scaling limit of this model is the
(1,2)-perturbation of $c=\tfrac{1}{2}$ conformal field theory due to
Zamolodchikov \cite{Za} which has E$_8$ symmetry.
Among the hierarchy of models, in their four regimes, which form the dilute
A$_L$ model \cite{wns}, is a realisation of this Ising model.

The A$_L$ model is an $L$ state interaction-round-a-face model \cite{bax2}
whose
adjacency diagram is the Dynkin diagram of A$_L$ with the additional
possibility
of a state being adjacent to itself on the lattice. In regime 2, the central
charge is
\begin{equation}
c=1-\frac{6}{L(L+1)},\label{cc}
\end{equation}
and for $L$ odd the $Z_2$ symmetry is broken away from criticality. As
well as general calculations for dilute A$_L$ with $L$ odd, various
results have been obtained which demonstrate the Ising critical exponents
\cite{wpsn}, \cite{BFZ} and hidden E$_8$ structures 
\cite{warp}-\cite{jimbo}
in the dilute A$_3$ model. Further, for $L=4$ and $L=6$, 
the central charge of the E$_7$ and E$_6$ field theories \cite{fz}
are recovered from (\ref{cc}). While no complete calculation of order parameters for dilute
A$_L$ with $L$ even has yet been carried out, there is a growing literature
concerning the hidden E-type structures \cite{warp}, \cite{BSc}, \cite{suz2}.

The motivation for the result presented here is a recent paper
\cite{ch} in which are given arguments for the  higher-order
terms in the scaling forms for the Ising free energy and mass spectrum, and
numerical estimates for some of the corresponding amplitudes and univeral
amplitude ratios. In Section \ref{review}  we review our previous results
for the
eigenspectrum of the dilute A$_L$ model for $L=3,4,6$, or equivalently the mass
spectrum for E$_8$, E$_7$ and E$_6$, with particular reference to the way the
inversion relation of the solvable model is expressed through it. In Section
\ref{higher} we calculate the first correction term for all three cases,
E$_{6, 7, 8}$ and, as a consequence of this property of the lattice model
realisation, we are able to give the coefficient very simply.

\section{The mass spectra} \label{review}
The eigenvalues of the row-to-row transfer matrix of the dilute A models
are \cite{BNW}
\begin{eqnarray}
\Lambda(u) &=& \omega \left[
\frac{\te(2\lambda-u)\;\te(3\lambda-u)}{\te(2\lambda)\;\te(3\lambda)}
\right]^N
\prod_{j=1}^N
\frac{\te(u-u_j+\lambda)}{\te(u-u_j-\lambda)}
\nonumber \\
&+& \left[
\frac{\te(u)\;\te(3\lambda-u)}{\te(2\lambda)\;\te(3\lambda)}
\right]^N
\prod_{j=1}^N
\frac{\te(u-u_j) \; \te(u-u_j-3\lambda)}
     {\te(u-u_j-\lambda) \; \te(u-u_j-2 \lambda)}
\label{eigs}
\\
&+& \omega^{-1}
\left[
\frac{\te(u)\;\te(\lambda-u)}{\te(2\lambda)\;\te(3\lambda)}
\right]^N
\prod_{j=1}^N
\frac{\te(u-u_j-4\lambda)}{\te(u-u_j-2\lambda)} \nonumber
\end{eqnarray}
where the $N$ roots $u_j$ are given by the Bethe equations
\begin{equation}
\omega \left[
\frac{\te(\lambda-u_j)}{\te(\lambda+u_j)}\right]^{N} =
-\prod_{k=1}^{N}
\frac{\te(u_j - u_k - 2\lambda) \; \te(u_j - u_k + \lambda) }
     {\te(u_j - u_k + 2\lambda) \; \te(u_j - u_k - \lambda) }
\label{BAE}
\end{equation}
and $\omega=\exp(\ib \pi \ell/(L+1))$ for $\ell=1,\ldots,L$. Here the
(standard) elliptic functions have nome $p=\e^{-\epsilon}$ and for the
regime of
interest, the spectral parameter obeys $0<u<3 \lambda$, where the crossing
parameter is $\lambda=\frac{\pi s}{r}$ and in terms of $L$,  $s=L+2$ and
$r=4(L+1)$.

Based on numerical data for the string
structure and positions of the Bethe ansatz roots for the dilute $A_3$ model
\cite{BNW}, \cite{GN}, which indicate that there are eight excitations
of thermodynamic significance, recurrence relations which enable the eigenvalue
expressions to be found were solved \cite {bs},
\cite{BSb}. The technique used is an exact perturbative approach, about the
ordered limit (with
$N$ large), first developed by Baxter for the eight-vertex model
\cite{bax1}, \cite{bax2}, and also applied to the
cyclic
solid-on-solid model \cite{pb}.
In terms of the conjugate variables
$w=\e^{-2 \pi u/\epsilon}$ and $x=\e^{-\pi^2/r\epsilon}$, the ordered limit is
$x\to 0$ ($w$ fixed). For the $L=3$ case, this corresponds to
the strong-field limit, while for the $L$ even cases, this
is the low-temperature situation. The elliptic functions in the conjugate
modulus
form of (\ref{eigs}) are those defined below in (\ref{ell}), with nome
$x^{2r}$.

The quantities actually determined are the excitations
\begin{equation}
r_j(w)=\lim_{N\to\infty}
\frac{\Lambda_j(w)}{\Lambda_0(w)},
\end{equation}
in terms of which the correlation lengths are $\xi_j^{-1} = - \log r_j$.
Because of the inversion and crossing relations obeyed by the model's Boltzmann
weights \cite{wpsn}, the excitations must satisfy the inversion relation:
\begin{equation}
r_j(w)
\, r_j(x^{6s}w)=1,\label{weak}\end{equation}
while a consequence of (\ref{eigs}) is the stronger relation, which implies the
former,
\begin{equation}r_j(w) \,
r_j(x^{4s}w)=r_j(x^{2s}w).\label{strong}\end{equation}

In terms of the (conjugate modulus) elliptic functions
\begin{equation}
E(z,q)=\prod_{n=1}^{\infty}(1-q^{n-1}z)(1-q^n/z)(1-q^n)=E(q/z,q)
=-zE(z^{-1},q),
\label{ell}
\end{equation}
the expression obtained \cite{BSb} is
\begin{equation}
r_j(w) = w^{n(a)} \prod_a \frac{E(-x^a/w,\,x^{60}) E(-x^{30-a}/ w,\,x^{60})}
                                {E(-x^a w,\,x^{60})
E(-x^{30-a}w,\,x^{60})}. \label{exc}
\end{equation}

Apparent in this expression
is the band structure of the eigenspectrum, labelled by powers of $w$, the
power given by the number of integers $n(a)$ appearing in the product.  The
values $a$ takes, arising from the calculation described above, are those given
in Table
\ref{eight}. Transforming to the original variables, appropriate for the
critical limit, the leading term is
\begin{equation}
m_j =\xi_j^{-1}\sim 8\, p^{8/15}  \sum_a \sin \frac{a\pi}{30}
\quad \mbox{as} \quad p \to 0, \label{mass}
\end{equation}
which gives the mass ratios of the E$_8$ field theory \cite{Za}.

There is no explicit $L$ dependence in the integers $a$ which could be
generalised to other members of the dilute A hierarchy in an immediate way.
However, they have appeared in connection with E$_8$ in various contexts,
for example
\cite{K}, and with various interpretations. Most significantly, they
occur in affine Toda theory, where they appear in the $S$-matrix for scattering
from the particle labelled $m_1$
\cite{BCD}. In the context of the dilute A$_3$ model, McCoy and Orrick
\cite{MO}
observed them in work related to \cite{BNW}, and hence to the same
Bethe ansatz root string structure used in the calculation of (\ref{exc}).
Suzuki \cite{suz} has used the quantum transfer matrix (QTM) approach to
recover
the E$_8$ Bethe ansatz equation from dilute A$_3$ \emph{without} any assumption
of particular string structure, and has remarked the occurence of the same
integers in the zeroes of the fusion QTMs. This
suggests that for the cases $L=4,6$ an expression analogous to (\ref{exc})
should
describe the eigenspectra, since the integers of Table \ref{eight} have
E$_6$ and E$_7$ counterparts.

In expression (\ref{exc}) `30' plays a distinguished role in relationship to
the new nome $x^{12s}$, where $s=L+2$. On the other hand, we see that in
(\ref{mass}) it enters as the dual Coxeter number $g=30$ of E$_8$.
Moreover, the
universal amplitude
\cite{F}, \cite{DM}
\begin{equation}f_s\xi_1^{2}=0.061728\ldots\end{equation}
is obtained \cite{bs}, and this relies on the power of $p$ in the correlation
length $\xi_j$ being appropriately related to that of the singular part of 
the free energy of the dilute  A$_L$ model \cite{wpsn},
\begin{equation}
f_s\sim  p^{r/3s} \quad \mbox{as} \quad p \to 0.
\end{equation}

Although (\ref{exc}) would obey (\ref{weak}) for any integer $a$,
the stronger inversion relation (\ref{strong}) is satisfied because each
integer
$a$ in Table
\ref{eight} occurs together with
$a+2s=a+\frac{g}{3}$, or equivalently, from properties of the elliptic function
(\ref{ell}), with $4s-a =\frac{2g}{3}-a$.

Gathering together these observations, it was proposed \cite{BSc} that
(\ref{exc}) is a special case of the expression
\begin{equation}r_j(w) = w^{n(a)} \prod_a \frac{E(-x^{\frac{6sa}{g}}/w,
x^{12s})
E(-x^{\frac{6s(g-a)}{g}}/ w, x^{12s})}
       {E(-x^{\frac{6sa}{g}}w, x^{12s})
E(-x^{\frac{6s(g-a)}{g}}w, x^{12s})}.\label{conj}
\end{equation}
For the dilute A$_4$ model, which is related to E$_7$, $g=18$, while for the
A$_6$ model the appropriate Coxeter number is that of E$_6$, $g=12$.

Insisting only that the integers $a$ appearing in (\ref{conj}) be
such that (\ref{strong}) is obeyed  and that they be E$_{6,7}$ analogues of
those
in Table \ref{eight} we were led to consider those in Table \ref{seven} and
\ref{six}. The mass ratios of E$_7$ and E$_6$ respectively are correctly given
by the leading term when
(\ref{conj}) is expressed in the original nome,
\begin{equation}
\xi_j^{-1}=m_j \sim 8\, p^{r/6s}  \sum_a \sin \frac{a\pi}{g}
\quad \mbox{as} \quad p \to 0.
\end{equation}

The integers in Tables \ref{seven} and \ref{six} correspond to scattering from
particle $m_2$ in the $S$-matrix of \cite{BCD}, unlike those in Table
\ref{eight}
which we noted before corresponded to $m_1$. This
appears to be related to the concrete connection drawn in \cite{HM} between
Toda
theory related to affine Lie algebras and integrable perturbations of conformal
field theory . The additional node on the Dynkin diagram in the affine case
connects to the node called $m_1$ (resp. $m_2$, $m_2$) in the field theory
labelling of the E$_8$ (resp. E$_7$, E$_6$) diagram, thus distinguishing this
node. We should also remark that Suzuki \cite{suz2} has recently obtained these
integers in analogous work to \cite{suz} for  $L=4, 6$.

The conjecture (\ref{conj}) has now been confirmed \cite{BSd} by using the
string structure for the Bethe ansatz roots for dilute $A_4$ \cite{GNtbp}
to perform the same type of calculation that led to (\ref{exc}) for the $L=4$ case.

\section{Higher-order terms}\label{higher}
 Recently, Caselle and Hasenbusch \cite{ch} have presented arguments based on
the renormalization group approach to give higher-order terms appearing in
the scaling form of the free energy and mass spectrum of the critical two-dimensional 
Ising model in a magnetic field, and have obtained numerical results for some
critical
amplitudes of these subleading corrections.
Since the dilute A$_3$ model is from the same universality class, it seems
pertinent to examine the higher-order terms in the mass spectrum (\ref{conj}),
to establish both their order, and the coefficients (from which universal
amplitude ratios can be constructed).

In \cite{ch}, the mass spectrum is given in the form (our notation for
coefficients)
\begin{equation}
m_j^2(h)={\cal A}_{m_j}^2 h^{\frac{16}{15}}\left( 1+{\cal
B}_{m_j}h^{\frac{16}{15}} +{\cal C}_{m_j}h^{\frac{22}{15}}
+{\cal D}_{m_j}h^{\frac{30}{15}}+
{\cal E}_{m_j}h^{\frac{32}{15}}+\ldots \right), \label{spec}
\end{equation}
which includes contributions from both relevant and irrelevant operators.

Now consider the dilute A expression (\ref{conj}) expressed in terms of
the original nome, $p$, which plays the role of the magnetic
field $h$ for $L=3$ (while for $L$ even $p=0$ corresponds to critical
temperature):
\begin{equation}
m_j = 2 \sum_a \log \frac{
\tv(\frac{a\pi}{2g}+\frac{\pi}{4},p^{r/6s})}
{\tv(\frac{a\pi}{2g}-\frac{\pi}{4},p^{r/6s})}. \label{expand}
\end{equation}
The solvable model can be expected to agree only with terms in (\ref{spec})
due to relevant operators.
Using the definition
\begin{equation}
\vartheta_4(u,q)=\prod_{n=1}^{\infty}\left(
 1-2q^{2n-1}\cos2u+q^{4n-2}\right)\left(1-q^{2n} \right)
\end{equation}
and the standard expansion of $\log(1+z)$, it is easy to see that
only odd-integer powers of $p^{\frac{r}{6s}}$ will occur, i.e.
$p^{\frac{16}{15}}$
for $L=3$ case. At first sight it is disappointing that the first higher-order
term arising from the dilute A$_3$ representation of the Ising model is
not immediately comparable with
the results of \cite{ch}, since their numerics do not extend to the fifth term in
(\ref{spec}).

Nevertheless, the coefficient of the term is interesting in its own right.
Expanding (\ref{expand})
\begin{equation}
m_j =8\, p^{\frac{r}{6s}}\left\{  \sum_a \sin \frac{a\pi}{g}
+\frac{4}{3}(p^{\frac{r}{6s}})^2 \sum_a \sin^3
\frac{a\pi}{g}+\ldots\right\},
\end{equation}
and taking out the coefficient of the
leading-order term (the mass amplitude) we are left to consider
\begin{equation}
\frac{\sum_a \sin^3 \frac{a\pi}{g}}{\sum_a \sin \frac{a\pi}{g}}
\end{equation}
which appears unwieldy, particularly since each set of possible $a$'s
may contain a different number of values.
However, the property of the sets of integers appearing in Tables
\ref{eight}-\ref{six}
which caused (\ref{conj}) to obey the inversion relation (\ref{strong}),
namely that $a$ occurs together with $a+\frac{g}{3}$ (or
$\frac{2g}{3}-a$),
means that by applying the simple trigonometric identity
\begin{equation}
\sin 3z=3 \sin z-4\sin^3 z,
\end{equation}
we obtain
\begin{equation}
m_j =8\, p^{\frac{r}{6s}}\left\{\sum_a \sin \frac{a\pi}{g}\right\}\left\{ 1
+(p^{\frac{r}{6s}})^2+{\mathcal O}((p^{\frac{r}{6s}})^4)\right\}.
\end{equation}

To summarize, a consequence of the inversion relation of the dilute A$_L$ model
is that in the mass spectra, as obtained from the solvable lattice
realisation, the next-to-leading correction term has the same simple
coefficient in each of the cases E$_{6,7,8}$. It will be interesting to
see if this value can be observed numerically for the Ising model in a
magnetic field.

\section*{Acknowledgements} This research is supported by an Australian
Research
Council small grant. The authors are grateful to Uwe Grimm and Bernard Nienhuis
for making available their unpublished numerical results.  They
acknowledge interesting and useful discussions with Junji Suzuki, Bernard
Nienhuis, Tomoki Nakanashi, Ole Warnaar and  Will Orrick.

\newpage


\newpage
\begin{table}
\begin{center}
\begin{tabular}{|c|c|l|}
\hline
$j$ &  $n(a)$& $a$ \\
\hline
1   &       2 & 1, 11 \\
2   &       2 & 7, 13 \\
3   &       3 & 2, 10, 12 \\
4   &       3 & 6, 10, 14 \\
5   &       4 & 3, 9, 11, 13 \\
6   &       4 & 6, 8, 12, 14\\
7   &       5 & 4, 8, 10, 12, 14\\
8   &       6 & 5, 7, 9, 11, 13, 15 \\
\hline
\end{tabular}
\caption{The integers $a$ which appear in the eigenvalue expression (\ref{exc})
of the dilute A$_3$ model.}
\label{eight}
\end{center}
\end{table} 

\begin{table}
\begin{center}
\begin{tabular}{|c|c|l|}
\hline
$j$ &  $n(a)$ & $a$ \\
\hline
1&      1 & 6 \\
2&     2  & 1, 7 \\
3&   2    & 4, 8 \\
4&    2   & 5, 7 \\
5&   3    & 2, 6, 8 \\
6&   3    & 4, 6, 8 \\
7& 4      & 3, 5, 7, 9 \\
\hline
\end{tabular}
\caption{The integers to appear in (\ref{conj}) in the $L=4$, or
equivalently E$_7$ case.}\label{seven}
\end{center}
\end{table}

\begin{table}
\begin{center}
\begin{tabular}{|c|c|l|}
\hline
$j$ & $n(a)$ & $a$ \\
\hline
1 ,$\bar 1$&      1 & 4 \\
2&     2  & 1, 5 \\
3, $\bar 3$&   2    & 3, 5 \\
4&    3   & 2, 4, 6 \\
\hline
\end{tabular}
\caption{The integers to appear in (\ref{conj}) in the $L=6$ case.}\label{six}
\end{center}
\end{table}

\end{document}